\documentclass[conference]{IEEEtran}
\IEEEoverridecommandlockouts
\usepackage{amsmath,amssymb,amsfonts}
\usepackage{algorithmic}
\usepackage{graphicx}
\usepackage{textcomp}
\usepackage{xcolor}
\usepackage{wrapfig}
\usepackage[font=small,skip=-5pt,textfont=bf]{caption}
\usepackage{url}
\usepackage{cite}

\def\BibTeX{{\rm B\kern-.05em{\sc i\kern-.025em b}\kern-.08em
    T\kern-.1667em\lower.7ex\hbox{E}\kern-.125emX}}

\makeatletter
\newcommand{\linebreakand}{%
  \end{@IEEEauthorhalign}
  \hfill\mbox{}\par
  \mbox{}\hfill\begin{@IEEEauthorhalign}
}
\makeatother

\author{\IEEEauthorblockN{Agrim Gupta\IEEEauthorrefmark{1},
Shenggang Dong\IEEEauthorrefmark{2},
Mehmet Mert Sahin\IEEEauthorrefmark{2},
Younghan Nam\IEEEauthorrefmark{2},
Frederik J. Harris\IEEEauthorrefmark{1} and
Dinesh Bharadia\IEEEauthorrefmark{1}}
\IEEEauthorblockA{\IEEEauthorrefmark{1}UC San Diego, \IEEEauthorrefmark{2}Samsung Research America}
}

\begin{document}
\title{Utilizing High Sampling-rate ADCs for Cost Efficient MIMO Radios\\
\thanks{Work done in part during an internship at Samsung Research America, Correspondence Email: agrim9gupta@gmail.com}
}

\maketitle

\begin{abstract}
In the past decade, $>1$ Gsps ADCs have become commonplace and are used in many modern 5G base station chips. 
A major driving force behind this adoption is the benefits of digital up/down-conversion and improved digital filtering.
Recent works have also advocated for utilizing this high sampling bandwidth to fit-in multiple MIMO streams, and reduce the number of ADCs required to build MIMO base-stations. 
This can potentially reduce the cost of Massive MIMO RUs, since ADCs are the most expensive electronics in the base-station radio chain.
However, these recent works do not model the necessary decimation filters that exist in the signal path of these high sampling rate ADCs.
We show in this short paper that because of the decimation filters, there can be introduction of cross-talks which can hinder the performance of these shared ADC interfaces.
We simulate the shared ADC interface with Matlab 5G toolbox for uplink MIMO, and show that these cross-talks can be mitigated by performing MMSE equalization atop the PUSCH estimated channels.
\end{abstract}

\section{Introduction}
\begin{figure}[t!]
    \centering
    \includegraphics[width=\linewidth]{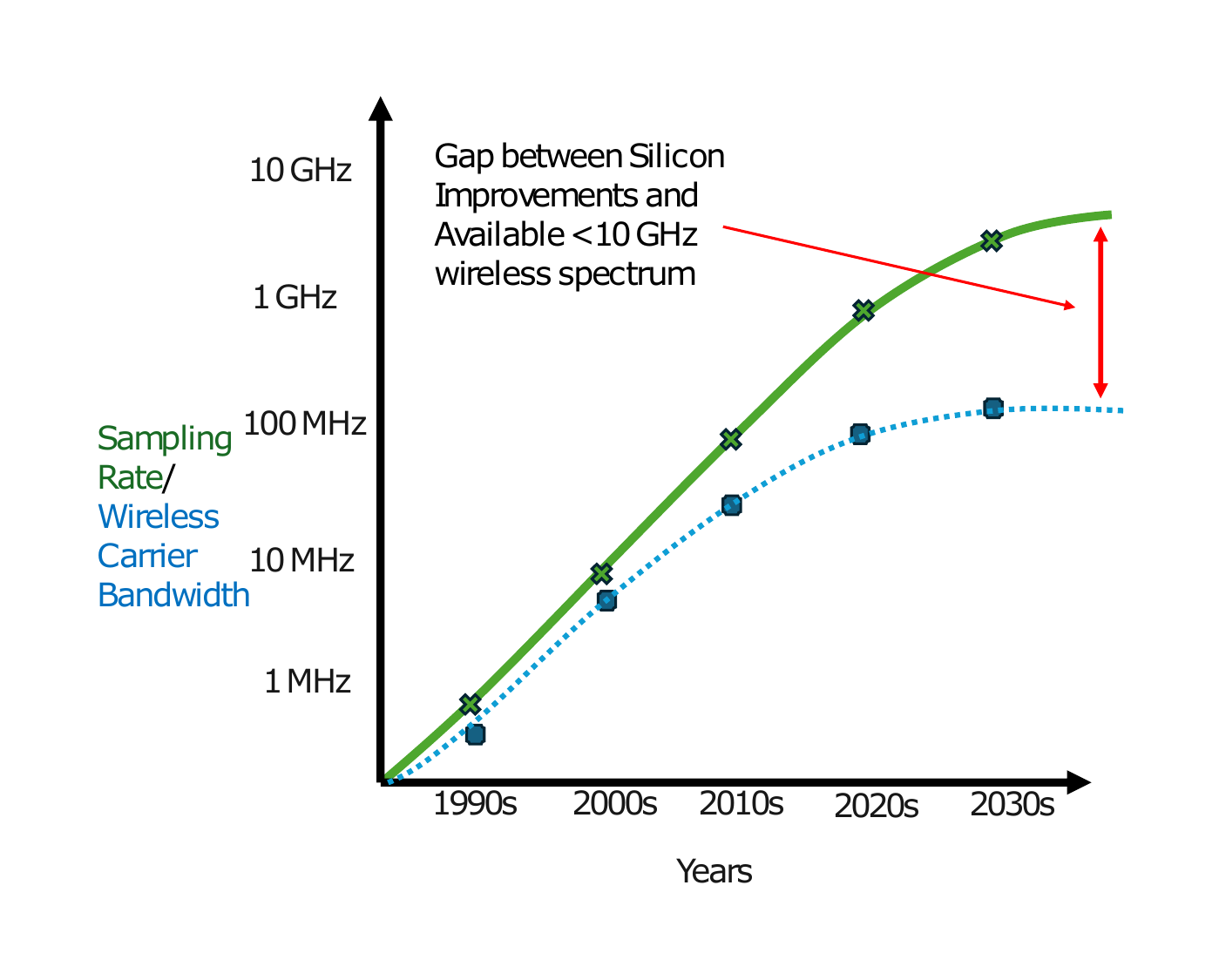}
    \caption{Over the past decades, sampling rates of dataconvertors have grown faster than wireless carrier bandwidths, which have saturated to a much lower number}
    \label{fig:ADC_trend}
\end{figure}

As the human technology prowess grows, and with better fabrication process, the sampling rates for dataconvertors (ADC/DAC) increase.
This allows faster data communications, both wired (optical links), and wireless (5G, Wi-Fi).
However, in case of wireless communications, especially 5G, utilizing these increased sampling rates for faster data communications is not as straightforward, since the speeds are dictated by the available over-the-air bandwidth, or spectrum.
That is, as communication technologies have hit ubiquitous deployments with 5G, the available wireless bandwidths in the hotly contested sub-10 GHz bands, have saturated to about 100-500 MHz \cite{qi2016quantifying,mihovska2020overview,mahon20175g}, whereas today the ADCs are capable of even supporting 10 GHz bandwidth \cite{gomez2016theoretical,kumar2023750mw,swindlehurst20218,heidari2025phasemo}, as shown in Fig. \ref{fig:ADC_trend}.

Hence, instead of supporting the higher rates, wireless communications use these high-sampling-rate ADCs for different features.
For example, typical 5G base-station chips have $\sim 5$ GHz sampling rate ADCs just to receive $\sim 100$ MHz signals, hence not fully utilizing the large rates available.
Instead, this large rate of oversampling allow for features like, (a) direct digital up/down conversion for the sub-6 GHz range of frequencies supported by modern RF SoCs \cite{chu2024integration,di2020rf}, (b) ability to digitally filter out interfering signals \cite{farley2018all,steiner20231} and (c) increased SNR due to lower noise-folding effects \cite{pei2023qtt,tuthill2022wide,jia2023research}.
Recent work, GreenMO \cite{gupta2023greenmo} shows how these oversampled ADCs can be used to fit multiple narrow-band signals from multiple antennas (MIMO), and hence lead to an increased utilization of the ADC's wider bandwidth and thus increased data-rates by combining multiple antenna streams.
Essentially, GreenMO \cite{gupta2023greenmo} interfaces $M$ different antennas, receiving $B$ bandwidth signals each, with a single $F_s = MB$ bandwidth ADC. This is built via a `switched-combiner' interface which gates signals from one antenna at a time, corresponding to one sampling period of the ADC ($\frac{1}{MB}$). Hence, due to this interface, the ADC samples cycle across the $M$ antennas across $M$ different $\frac{1}{MB}$ sample times, and thus each antenna sample is captured across the net $\frac{1}{B}$ time period, which guarantees perfect reconstruction as per the Nyquist theorem.
The authors build a system prototype and showcase multiplexing $M=4$ streams of $B=20$ MHz streams across a single $MB=80$ MHz ADC. However, in order to be compatible with 5G bandwidths of $B=100$ MHz, $M=4/8$ antenna ports, 5G base-station ADCs which can have sampling rates $F_s = 5$ GHz, the techniques discussed in previous work should be further generalized to $F_s > MB$.

\begin{figure}[t!]
    \centering
    \includegraphics[width=0.6\linewidth]{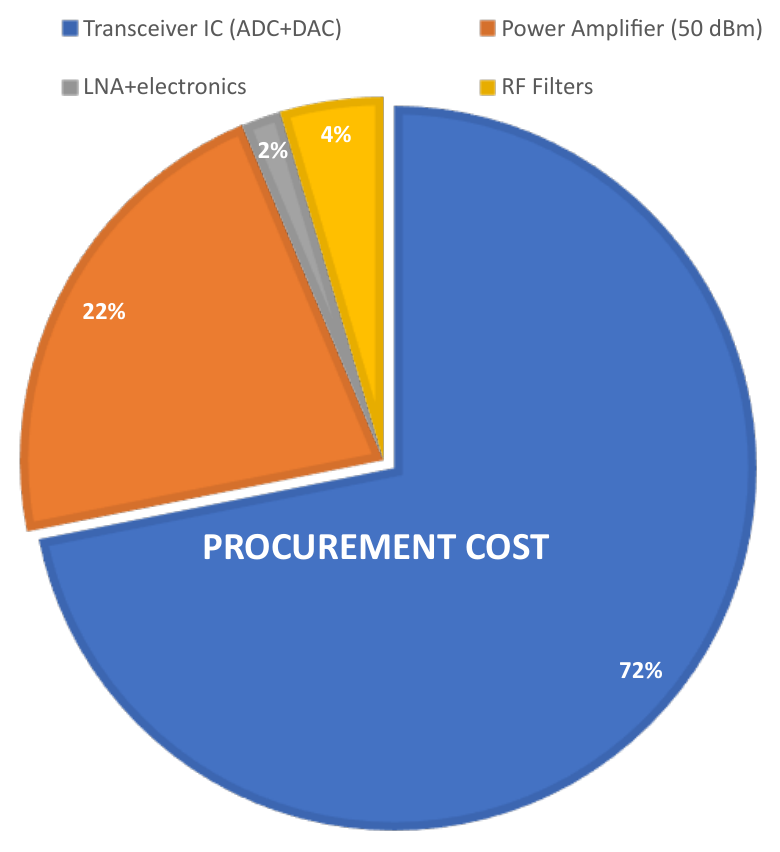}
    \vspace{10pt}
    \caption{Transceiver ICs (like ADRV9026 \cite{ADRV9026}) which consist of the ADC/DAC interfaces, end up being the major cost component, contributing to $>70\%$ of the total cost of a 5G base-station radio}
    \label{fig:pie}
\end{figure}

\begin{figure}[t!]
    \centering
    \includegraphics[width=\linewidth]{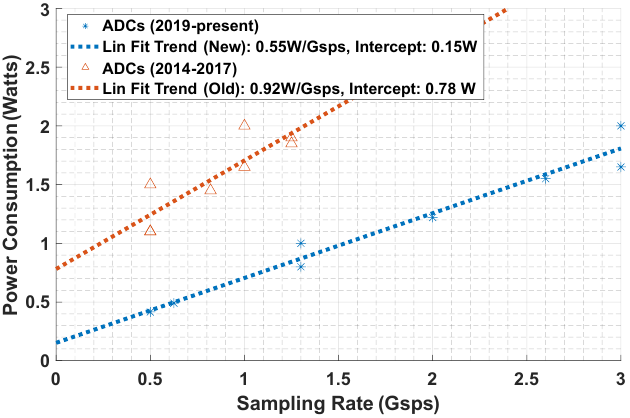}
    \vspace{3pt}
    \caption{Due to changes in process, faster sampling rate ADCs become more energy efficient. This plot considers power consumption of various sampling rate, 14 bit ADCs, with SNDR$>50$ dB released by Analog Devices from 2014 - 2024}
    \label{fig:adc_trend}
\end{figure}

The challenges posed by the said $F_s > MB$ generalization is due to presence of a decimation filter which would ultimately decimate from $F_s \to MB$. This decimation operation hinders the orthogonality between the time-gated per-antenna samples, and introduces cross-talks between multiple antennas sharing the same ADC. 
In this paper, we first generalize the mathematical operations in prior work to $F_s>MB$, show how to model the decimation filter introduced cross-talk mathematically, and via simulation based on Matlab 5G toolbox, we also show how MMSE based MIMO equalization atop channels estimated via PUSCH reference signals can resolve this cross-talk and recover the SNR to original levels. 

This generalization to GreenMO's antenna multiplexing to $F_s > MB$ allows $M$ antennas to share just a single digital port of commercially available transceiver ICs like ADRV9026 \cite{ADRV9026}, which satisfy such constraints.
This can help reduce the CapEx requirements of base-station radio-units (RUs), since transceiver ICs are the most expensive electronics in a typical RU, as shown in Fig. \ref{fig:pie}. 
Furthermore, improvements in process technology, further advocate such faster $F_s >MB$ sampling rates from reduced energy consumption viewpoint, as can be seen in Fig. \ref{fig:adc_trend}.
Lower cost, and energy-efficient RUs has also been touted to be an important direction towards a successful transition between 5G to 6G in the next few years \cite{rappaport2024waste, andrews20246}, and hence further explorations in such shared multiple antenna interfaces for MIMO RUs is pivotal to the next generation of wireless communications. 

The rest of the paper is organized as following. Section 2 discusses mathematical system model for MIMO receivers for $B$ bandwidth signals, both for traditional MIMO systems having $M$ $F_s =B$ ADCs for $M$ antennas, as well for past work considering a single $F_s = MB$ ADC to interface $M$ antennas. Section 3 generalizes the same mathematical model for higher sampling rates $F_s > MB$, which creates cross-talks, and then shows how such cross-talks can get absorbed in the wireless channel itself, and hence can be mitigated via MIMO equalization. Section 4 presents the simulation results using Matlab 5G toolbox, and Section 5 concludes with a summary of contributions and future work.

\section{System Model}
\label{sec:sysmodel}

In this section, we first describe the typical mathematical model for a $M$ antenna MIMO receiver front end, consisting of $M$ antennas, having $M$ RF chains each, to receive a $B$ bandwidth signal. 
Then, we describe the traditional ADC interface with $M$ different ADCs sampling at $F_s$, with $F_s>B$, to digitally sample the $B$ bandwidth signal.
Building upon, we describe the single ADC interface proposed in past work which sets $F_s = MB$, to reconstruct the $M$ different $B$ bandwidth signals from each antenna. 
However, it may not be always possible for ADC sampling rate to be fixed $F_s=MB$, and hence, we will generalize the mathematical model to $F_s > MB$ to build upon the previous work, as well, highlight the new challenges posed by this change.

\subsection{$M$ antenna MIMO receiver front end modelling}
We can represent the received signal at $i$-th antenna via the following Equation \eqref{eqn:antenna_rx} 
\begin{equation}
x^A_i(t)  = a^B_i(t)e^{j2\pi f_c t}   + w^a_i(t)
\label{eqn:antenna_rx}
\end{equation} 
where $a^B_i(t)$ is a bandlimited signal, with bandwidth $B$ loaded to a carrier frequency $f_c$, In addition to the signal, there is analog white noise at the antenna as $w^a_i(t)$ depending upon the antenna's temperature. The SNR at this stage can be written as ${SNR}^A = 10\log_{10}(\frac{||a^B||^2}{||w^A||^2})$
MIMO front end designs require $M$ different RF chains to filter, amplify and downconvert the received signal.
For simplicity, we can describe the signal after RF chain as $R(.)$ function, with the output signal $x^R_i(t) = R(x^A_i(t))$.
The $R(.)$ function is implemented by appropriate analog gain control (AGC) of LNA, downconversion mixer and RF filters, and these electronics are linear across the bandwidth $B$ to ensure $x_i(t) \approx Ga_i(t)+w_i(t)$, where $G$ is the RF chain gain and $w^R_i(t)$ is the noise term after the RF chain. Hence, Equation \eqref{eqn:rfc_rx} represents the analog signal after RF chain:
\begin{equation}
    x_i(t) = R(x^a_i(t)) \approx Ga_i(t)+w_i(t) 
    \label{eqn:rfc_rx}
\end{equation}
The Noise Figure (NF) is defined as the ratio (in dB) of the noise power before/after RF chain $NF = 10\log_{10}(\frac{||w_i||^2}{||w^A_i||^2})$, with SNR at front end interface output reducing accordingly to be $SNR^{FE} = SNR^{A}-NF$.

\subsection{Digital Interface with $M$ separate ADCs}
After the front-end interface takes the antenna signal $x^A_i(t)$, downconverts and amplifies it to $x^R_i(t)$, the designed digital interface aim is to obtain digital samples $A_i[n]$ which represent the sampled version of the received analog signal $a^B(t)$. 
Typically the gain term $G$ in front end interface is chosen such that the analog input to the ADC is full scale, that is $||x_i(t)|| \approx 1$, so that the quantization noise is minimized.

Next, we describe how the amplified and downconverted signal is digitized by the ADC interface. 
The $i$-th ADC takes $x_i(t)$ as input and produces a digitized version as output, $X_i[n] = x_i(nT_s)$, using a sample and hold process, with sampling time $T_s = \frac{1}{F_s}$, and   $F_s>B$. After the sample and hold process, a programmable decimation filter comverts the digitized signal to $B$ bandwidth, to make the downstream PHY processing optimum and not exacerbate the demands on digital interfacing.
Hence, given an input signal $x_i(t)$ to the ADC, the output digital signal from ADC is written as per Equation \eqref{eqn:adc_op}:
\begin{equation}
    X_i[n]=\mathcal{D}(x^R_i(nT_s))
    \label{eqn:adc_op}
\end{equation}
where $T_s = \frac{1}{F_s}$, $x_i(nT_s)$ represents the sampled value at $nT_s$, and $\mathcal{D}$ is a decimation filter which decimates $F_s \to B$. Typically, $F_s \sim 1-5$ GHz, $B=100-400$ MHz.
Since the analog signal is band-limited to $B$ bandwidth, the baseband representation of $a_i(t)$ is between $-\frac{B}{2},\frac{B}{2}$.
Hence, by sampling at $B$ rate (after decimation), $X_i[n]$ is successful in reconstructing the received signal $a_i(t)$, with a certain SNR, as shown in equation 
\begin{equation}
    X_i[n]=A_i[n]+W_i[n]
    \label{eqn:adc_op_wnoise}
\end{equation}
where $A_i[n]$ are $\frac{1}{B}$ separated samples of $a_i(t)$. 

The final output SNR is $\text{SNR}_{OP}(F_s) = 20*\log_{10}(\frac{||A_i||}{||W_i||})$. This output SNR depends on the ADC sampling rate, since the analog white noise term after LNA $W_i(t)$ folds back into the nyquist zone dictated by $F_s$ rate of ADC. 
If $F_s \to \infty$, $\text{SNR}_{OP} \to \text{SNR}_{FE}$, otherwise, there is a reduction in SNR such that $\text{SNR}_{OP} = \text{SNR}_{FE} - \Delta(F_s)$.
Hence, as $F_s$ is chosen such that $F_s >> B$, the SNR increases, since lesser noise folds back into the main signal band.


\subsection{Digital Interface with a single ADC, $F_s=MB$}

\begin{figure}
    \centering
    \includegraphics[width=\linewidth]{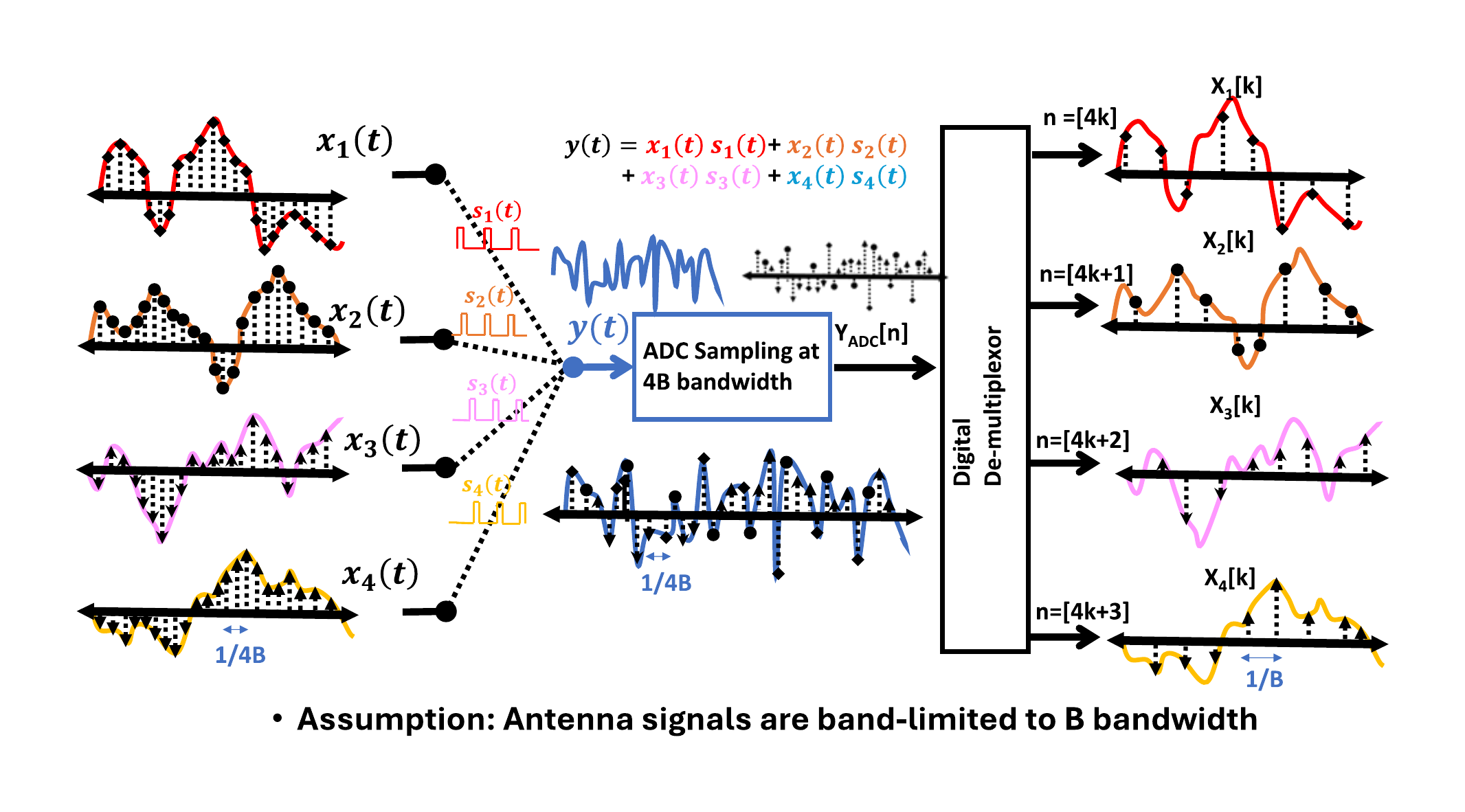}
    \caption{Reconstruction of original $B$ bandwidth antenna signal from a single $F_s=MB$ ADC ($M=4$ in the figure)}
    \label{fig:adc_recon_fig}
\end{figure}

In the past work, GreenMO \cite{gupta2023greenmo}, the authors show that the output SNR for $M$ antennas, sharing a single $F_s=MB$ bandwidth ADC is about the same as $M$ antennas using a $F_s=B$ bandwidth ADC. 
The main idea behind this is to introduce a `switched-combiner' interface, where the $M$ antennas signals are combined into a single interface by clocking each antenna's signals such that they occupy $\frac{1}{MB}$ time sample.
This is described intuitively in Fig. \ref{fig:adc_recon_fig}.

What happens here is that the $x_i(t)$ analog signals after the RF chain, go through the switched-combiner, such that the output from the interface is described in Eqn \eqref{eqn:switch_comb}

\begin{equation}
    y(t) = \sum_{i=1}^M x_i(t)c_i(t)
    \label{eqn:switch_comb}
\end{equation}

and the clocks $c_i(t)$ are $\frac{1}{M}$ duty cycled on-off codes, described in Eqn \eqref{eqn:duty_cycle}

\begin{equation}
    c_i(t) = \begin{cases}
                        1, \text{t $\in [\frac{i}{MB},\frac{i+1}{MB}]$} \\
                        0, \text{otherwise}
                    \end{cases}
    \label{eqn:duty_cycle}
\end{equation}

The design of the clocks ensure orthogonality in time domain, since only one of the clocks is $1$ at a given instant of time. 
That is, $\sum_t c_i c_j = 0, i \neq j$.
The output signal from this switched-combiner interface, with orthogonal clocks, is then sampled with a ADC with $F_s = MB$, to obtain $Y_{\text{ADC}}[n] = y(\frac{n}{MB})$. These $Y_{\text{ADC}}[n]$ samples are then de-interleaved to get back $x_i[k]$, where $n=Mk+i$. Here $x_i[k]$ denote the samples from $i$-th antenna, spaced $\frac{1}{B}$ apart, and hence denote the digitized samples $X_i[n]$ with $F_s = B$ bandwidth.
Hence, this switched-combiner scheme with $F_s = MB$ bandwidth ADC, reconstructs the $M$ $X_i$ signals with same SNR as the traditional scheme utilizing $M$ different ADCs with $F_s = B$ bandwidth ADCs.
This is because, time-domain orthogonality of clocks ensures that the de-interleaved signals $X_i[n]$ only come from antenna $i$'s RF chain going through $c_i$ time clock.

In the next section, we will detail and generalize this model discussed in GreenMO, to $F_s > MB$ sampling rates, which makes these orthogonal clocks to get afflicted by decimation filters, that breaks their time orthogality.

\section{Single ADC with $F_s > MB$ for $M$ antennas}

Next, we describe how we can build a shared ADC interface, where we show that a single ADC, which has $F_s > MB$. 
In particular, we explore the case where such $F_s>MB$ ADCs are interfaced in the digital domain only with $MB$ data-rate, and hence the signal needs to go through a $F_s \to MB$ decimation filter beforehand.
We will first show how the orthogonality between the codes break down due to the low-pass filtering required pre-decimation, and how this can be mitigated using MIMO equalization. Later, through the simulations based on Matlab 5G toolbox, we show that this mitigation recovers the original signals with no SNR loss.

\subsection{How low-pass filtering pre decimation breaks orthogonality}

\begin{figure}[t!]
    \centering
    \includegraphics[width=\linewidth]{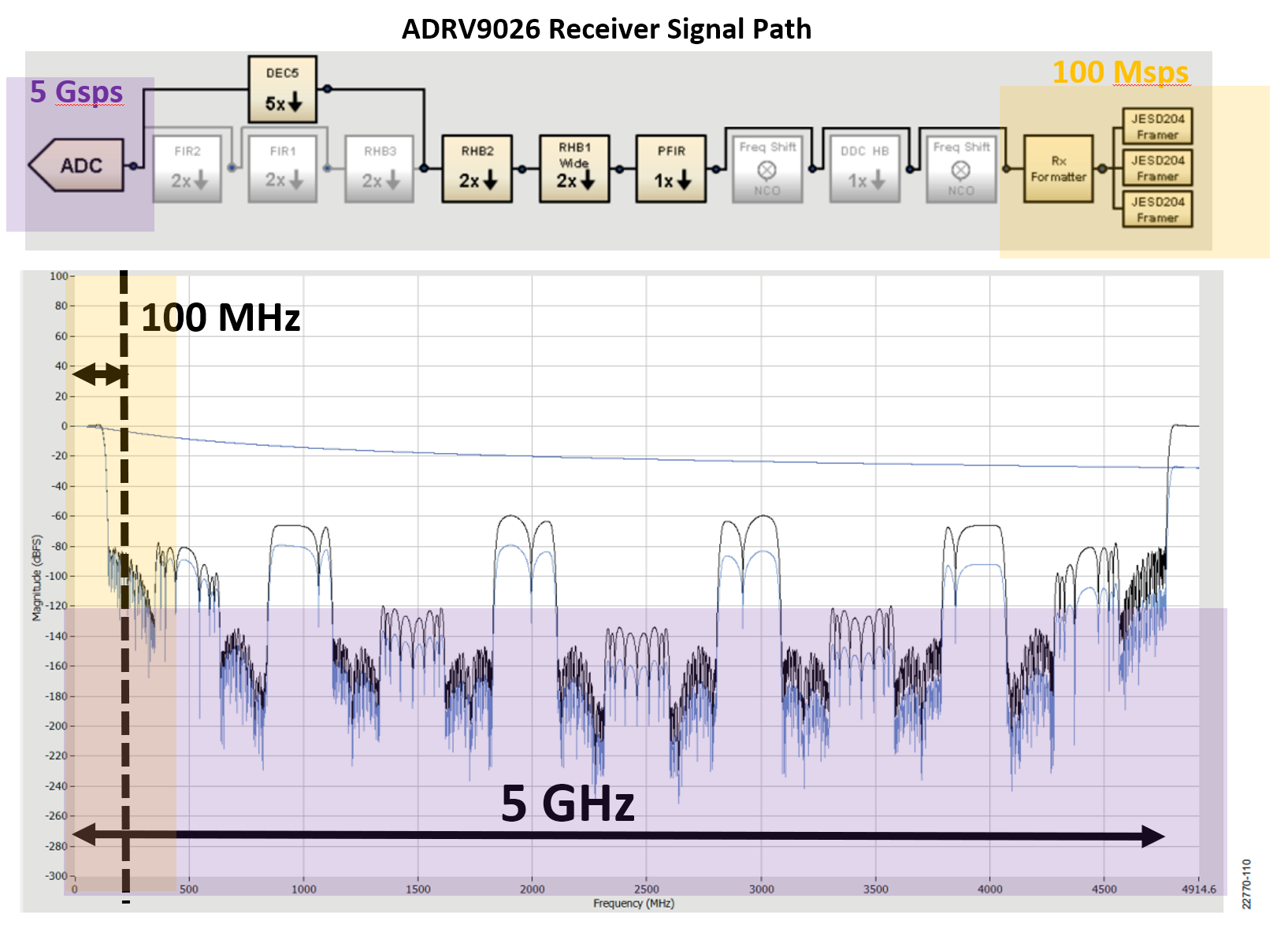}
    \caption{ADC interface of a popular 5G base-station transceiver IC ADRV9026}
    \label{fig:ADRV9026}
\end{figure}

\begin{figure}
    \centering
    \includegraphics[width=\linewidth]{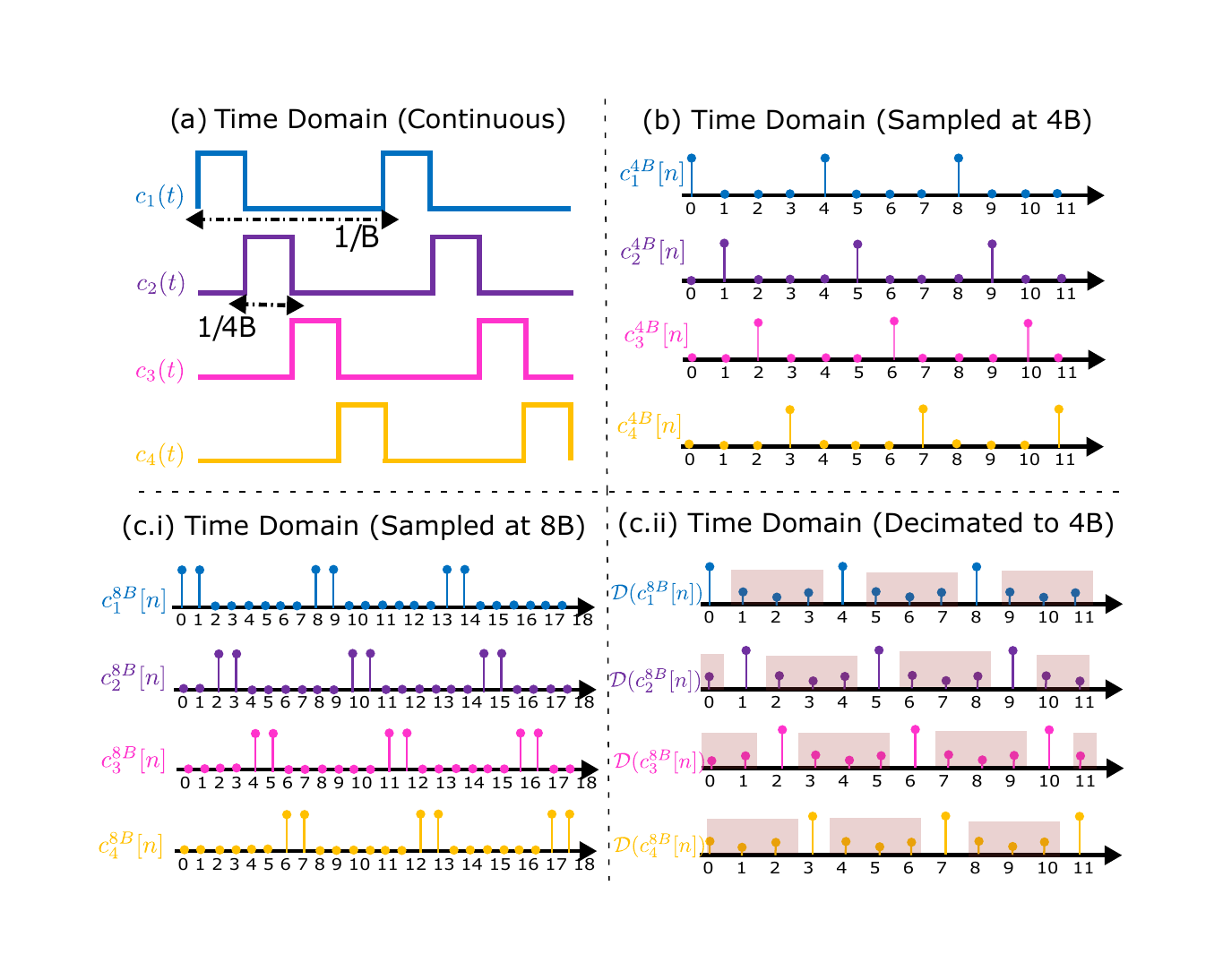}
    \caption{Clock designs decimate and original}
    \label{fig:clk_dig}
\end{figure}

\begin{figure}
    \centering
    \includegraphics[width=\linewidth]{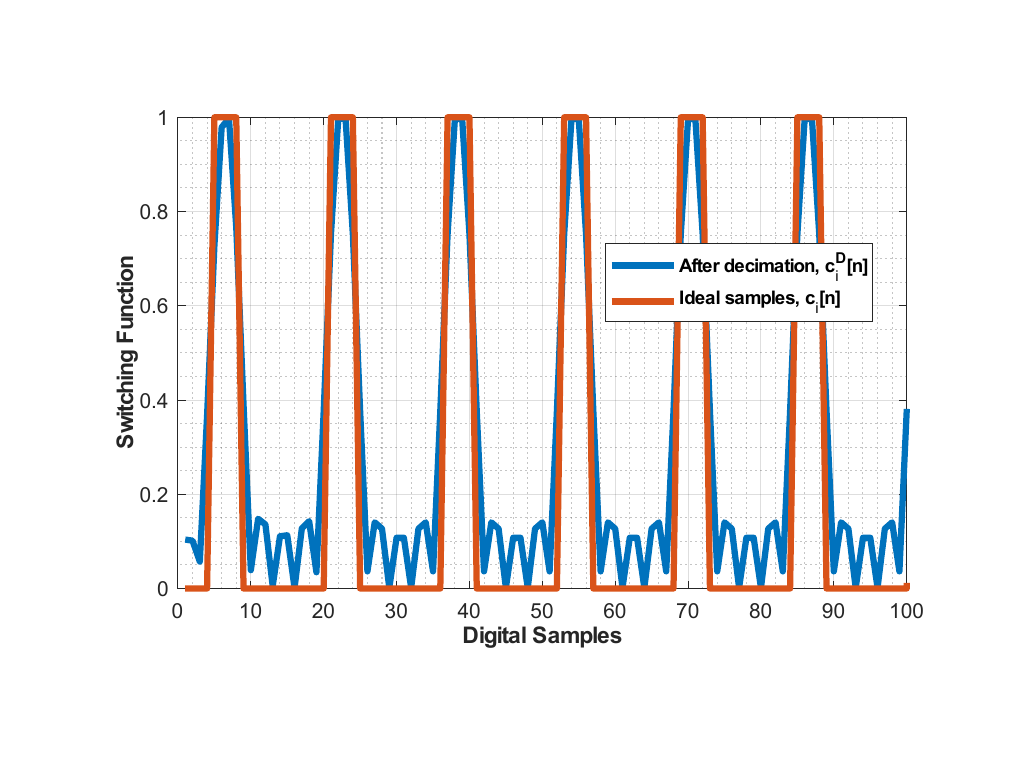}
    \caption{Decimation breaks down the orthogonality between codes}
    \label{fig:clk_decim}
\end{figure}

When we start generalizing this concept to $F_s >MB$, we require a decimation filter which will first use a digital low pass filter to reduce the signal from $F_s \to MB$, and then decimate it before sending it to the modem for higher layer processing. The reason modern ADCs have this oversampled+decimation construct is because the higher rate of sampling and the decimation later allows for lower noise folding and capabilities to handle the jammers via digital filtering. Because of the decimation filter later, the interface required between the high-sampling rate ADC and MIMO modem processor is also at the optimum lower rate, and hence this low-pass filter post ADC is a critical part of modern base-station receive chain. The flowchart describing internals of the ADC interface for a popular 5G base-station transceiver IC, ADRV9026 is shown in Fig. \ref{fig:ADRV9026}.

Intuitively, what happens because of the low-pass filtering after ADC sampling is that, it corrupts the orthogonality of the clocking scheme that allows for gating different antenna signals for different ADC sample. This is illustrated visually in Fig. \ref{fig:clk_dig}, where (a), (b) show the time domain clocks in continuous and digital $MB$ ($M = 4$ in Fig. \ref{fig:clk_dig}) domain. If the continuous domain clock is also sampled at $2MB>MB$, as shown in Fig. \ref{fig:clk_dig}(c), the orthogonality still remains, however, if the clock signal is decimated to $MB$, then the orthogonality vanishses since the low-pass filtering creates cross-talk among the codes, shown in red shade. This happens because the clock signal is a continuous signal, with frequency domain harmonics spreading beyond $MB$, and as shown in past work $MB$ digital sampling folds back the harmonics and preserves the orthogonality. But, presence of a low-pass filter breaks this assumption and throws away the extra harmonics, which leads to corruption of the clock signal. An example is shown in Fig. \ref{fig:clk_decim}, where the clock signal is sampled at $4MB$ and then low pass filtered at $MB$, $M=4$.

\subsection{Mathematical modelling, and addressing the cross-talk via MIMO equalization}

To model the loss of code-orthogonality mathematically, and understand why MIMO equalization will address this, we need look at start by modelling the shared ADC interface with $F_s > MB$. Following up from Section \ref{sec:sysmodel}, we have the analog signals after the RF chain for the $M$ antennas denoted as $x^R_i(t), i=1,2\ldots M$. We denote the switching codes per-antenna as $s_i(t)$ implemented by the RF switch. Hence, the input to the ADC, after the switched-combiner, denoted as $y(t)$ is given by the following equation
\begin{equation}
    y(t) = \sum_{i=1}^M x_i(t)s_i(t)
\end{equation}

The ADC would sample and hold to create a $Y_{SH}[n]$ at a rate of $T_s =\frac{1}{F_s}$, to have $Y_{SH}[n] = y(nT_s)$. Atop this, there will be a low pass filter of bandwidth $MB$, to create a $Y_{LPF}[n] = \mathcal{L}(Y_{SH}[n])$. Finally, there will be a $F_s \to MB$ decimation, with decimation ratio $D = \frac{F_s}{MB}$. Thus, the output of the $F_s>MB$ ADC, $Y_{ADC}[n_d]$ can be written via the following equation
\begin{equation}
    Y_{ADC}[n_M] = \downarrow D (L(Y_{SH}[n])) = \mathcal{D}(y(nT_s))
\end{equation}
where we represent $\mathcal{D}(.)$ as the composite decimation function absorbing the decimation operation, low pass filtering, as well sample and hold functionality. 

However, as a consequence, the low-pass filter throws away the harmonics beyond $MB$ frequency, which leads to two issues, (a) this leads to some loss in the signal amplitude since useful spreaded part of the signal is thrown away and (b) it breaks the orthogonality between the switching sequences $s_i(t)$. Hence, the reconstruction equations for shared ADC interface can be modeled as the following Equation \eqref{eq:math_lin}:

\begin{equation}
    \hat{x}_i[n] = a_i x_i[n]+\sum_{j\neq i}^M c_{ij}x_j[n]+w_i[n]
    \label{eq:math_lin}
\end{equation}

where $a_i<1$ represents the loss in signal due to the low-pass filtering, and $c_{ij}$ represents the cross talk components. Hence, we can write the Equation \eqref{eq:math_lin} in a matrix form as Equation \eqref{eqn:matrixform_expanded}:

\begin{equation}
\hat{X}[n] = 
C[n]X[n] + W[n]
\label{eqn:matrixform_expanded}
\end{equation}
\begin{equation}
    C[n] = \begin{bmatrix}
a_1 & c_{12} & c_{13} & \ldots & c_{1M}\\
c_{21} & a_{2} & c_{23} & \ldots & c_{2M}\\
c_{31} & c_{32} & a_{3} & \ldots & c_{3M}\\
.   & . & . & \ldots & . \\
.   & . & . & \ldots & . \\
c_{M1} & c_{M2} & c_{M3} & \ldots & a_{M}\\
\end{bmatrix}
\end{equation}
where $\hat{X}[n], X[n]$ are $M\times1$ vectors representing the reconstructed, and the original digital signal at $n$-th sample for $i=1,2\ldots M$ antennas, and $w[n]$ represents the noise terms per antenna.
$C[n]$ is a $M\times M$ matrix that represents the cross talk due to the decimation filters.
In $C[n]$, $a_i \approx 1$, and $a_i >> c_{ij}, i \neq j$ because of the tapering nature of the switching harmonics.
That is, the main harmonics are at $\pm B$, then we have $\pm 2B$ which is about $M$ dB lower, then $\pm 3B$ about $2M$ dB lower and so on. Hence, the harmonics beyond $MB$ are substantially weaker, which creates overall only about $0.1$ dB loss for typical $M=4$, and hence the cross talks by themselves aren't higher than the self term.
This guarantees that the $C[n]$ matrix is invertible (as the rows can be shown to be linearly independent due to the presence of a strong diagonal nature to the matrix). 
This allows addressing of the cross-talks by inverting the $C[n]$ matrix via traditional ZF/MMSE based MIMO equalization.

So far, we were considering the accuracy at which the per-antenna signals can get reconstructed as they go through the shared ADC interface. If instead, the goal is to directly perform MIMO processing, and decode the original user data for $M$ layers, which goes through wireless channel $M\times M$ channel $H$, then, the Equation \eqref{eqn:matrixform} can be written instead as 

\begin{equation}
\hat{X}[n] = 
(C[n]H[n])X[n] + W[n]
\label{eqn:matrixform}
\end{equation}

wherein, in order to decode the data, the required matrix to invert would now become $CH$ instead of $H$. Since we discussed that $C$ is invertible, it will not change the invertibility/condition number of the channel $H$, and hence the MIMO equalization block would end up inverting the composite channel matrix $CH$ instead of $H$. Hence simple ZF based MIMO equalization handles the cross-talks and gets back the SNR. The assumption of $C$ matrix invertibility breaks if the filtering cutoff frequency is lower than $MB$ and main harmonics are thrown away. Hence, for such sharper filters, there is a loss in SNR due to cross-talks. We show evaluation of reconstruction SNR for different low-pass filter cutoff frequencies in the next section. 

\begin{figure*}
    \centering
    \includegraphics[width=\linewidth]{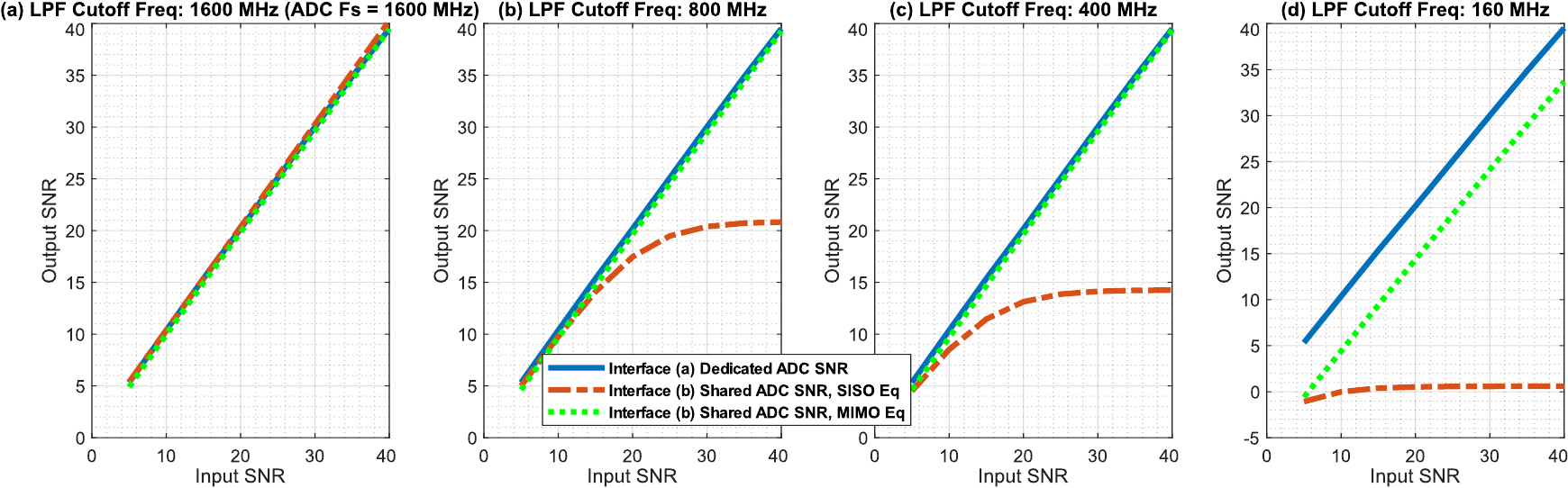}
    \vspace{3pt}
    \caption{We show the average reconstructed SNR across the $M=4$ antennas, $B=100$ MHz NR OFDM uplink signals, in (i) low pass filter is bypassed, ensuring orthogonality, and hence SISO equalization suffices to recover SNR, (ii)-(iii) Cutoff frequency is $>400$ MHz, which satisfies the conditions for MIMO equalization, and hence the ZF based equalization recovers the SNR to the expected levels, and hence no noise figure degradation, in (iv) there is a SNR penalty of $\sim7$ dB because of filtered out higher harmonics, leading to an ill formed matrix which even after MIMO equalization leads to a loss in SNR showing up as NF degradation}
    \label{fig:results1}
\end{figure*}
 
\section{Simulation Setting}
We utilize Matlab 5G toolbox for waveform creation, and utilize 5G compliant equalization process to show that cross-talk issue central to the generalization of past work to $F_s > MB$ ADCs is handled by simple ZF based MIMO equalization. The code used for result in this paper is hosted on a public github repository \cite{gitrepo}. Monte carlo simulation is performed with $5$ seeds, and SNRs averaged across 1 frame worth of data transmissions. The waveforms have $100$ MHz bandwidth and utilize $273$ RBs.

\begin{figure}
    \centering
    \includegraphics[width=\linewidth]{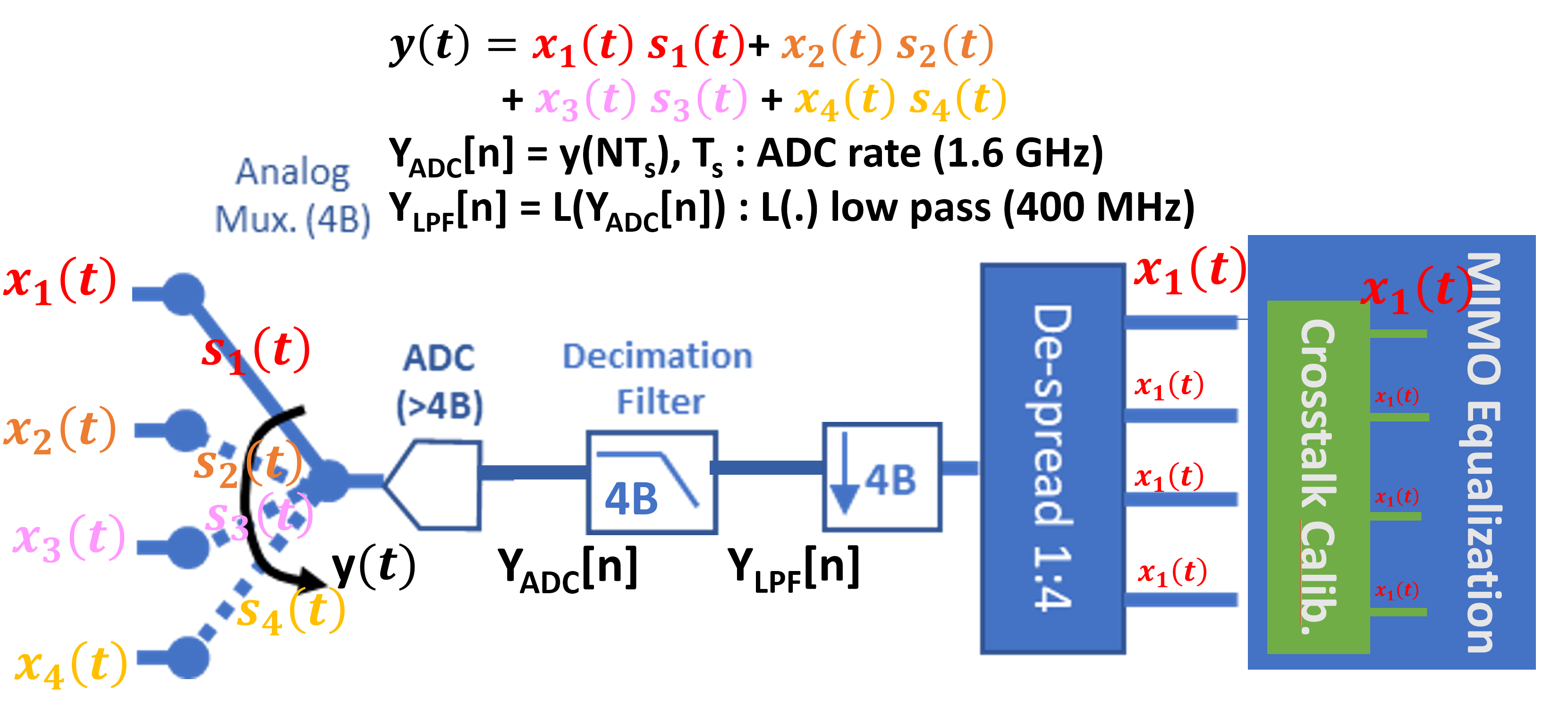}
    \caption{Visual illustration of the overall simulation setup}
    \label{fig:flowchart}
\end{figure}
\begin{figure}
    \centering
    \includegraphics[width=\linewidth]{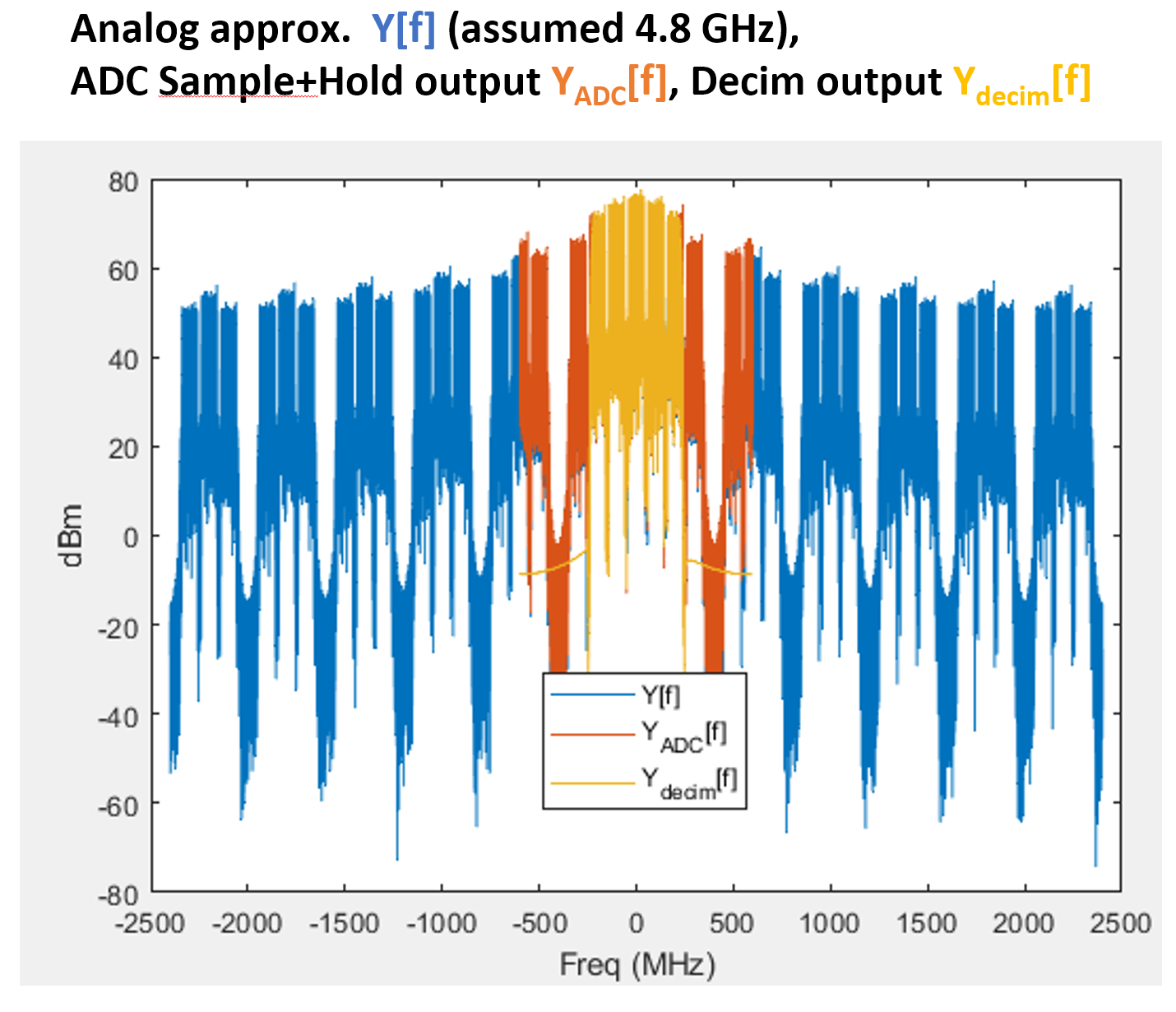}
    \caption{Spectrum Output showing frequency domain signals. As seen clearly, the low pass filtering throws away harmonics, hence corrupts the orthogonality}
    \label{fig:spectrum}
\end{figure}

\subsection{Waveform generation and filtering}
We create $4$ layer, $B=100$ MHz waveforms for uplink MIMO using Matlab 5G toolbox for our simulation purposes. 
The PUSCH constellation used in simulation is QPSK, since it allows us to accurately obtain SNR from the EVMs, for SNRs above $3$ dB, as QPSK is decoded properly for all those SNRs of interest.
For simplicity, we assume $4$ layers represent $4$ physical antenna ports (digital beamforming), however, this can be generalized easily to assume $4$ combined antenna ports that need to be multiplexed (hybrid beamforming).
These waveforms represent the signals after the RF chain amplification and filtering, and just before the digital interfaces.
Since these waveforms would actually be in analog domain, we upsample it by $160$x to create an approximation for the analog waveform. 
That is, we effectively treat analog waveform as digital waveform sampled at $16$ GHz for simulation purposes.

Atop these waveforms, we simulate various digital interfaces, including (a) the traditional individual ADCs per antenna, with $F_s = 4*100 = 400$ MHz, with $4x$ oversampling factor to improve SNR (b) single ADC with $F_s = 4*4*100 = 1600$ MHz to accommodate $4$ streams, as well get the SNR gains from the $4$x oversampling. 
In the downstream processing, the interface (a) will require a $100 MHz$ low-pass filter, and $4x$ decimation to match the bandwidth back to $100$ MHz. 
This allows improved SNR, since the SNR will correspond to noise level from $400$ MHz noise folding of LNA analog noise, while only having $100$ MHz worth of data to the downstream processor.
For downstream processing of interface (b), first there will be a decimation from $1600 \to 400$ MHz, including a low-pass filter of $F_c$ cutoff. Ideal value of $F_c=400$ MHz, but we will vary $F_c$ to show effects of low-pass filtering in our evaluations. After this low-pass filtering, the $4$ different $400$ MHz chunks are segregated from the antenna gating codes, and then there is another $400 \to 100$ MHz decimation, similar to interface (a) which finally gets the $4$ different 100 MHz signals, with similar SNR as interface (a). The goal of the simulation is to explore if there is any SNR penalty, imposed due to the extra low pass filtering at $F_c$ and decimation from $1600\to400$ MHz step, which separates the two interfaces. This is also explained via a flowchart figure in Fig. \ref{fig:flowchart}.

\subsection{Downstream Processing}

Once the $100$ MHz digitized waveforms are obtained via both the interfaces, they are fed to the MIMO processor code written with Matlab 5G toolbox as well. 
This code demodulates the NR-OFDM symbols, and performs practical channel estimation atop the PUSCH reference signal.
Then after obtaining the channel estimates, it utilizes MIMO equalization (MMSE) to obtain the equalized symbols, from which we calculate the EVM.
From the EVM, we obtain the SNR using $SNR = -20*\log_{10}(EVM/100\%)) + K$, where $K$ is a normalization constant that we tune to match the input SNR.
\begin{figure}[t!]
    \centering
    \includegraphics[width=\linewidth]{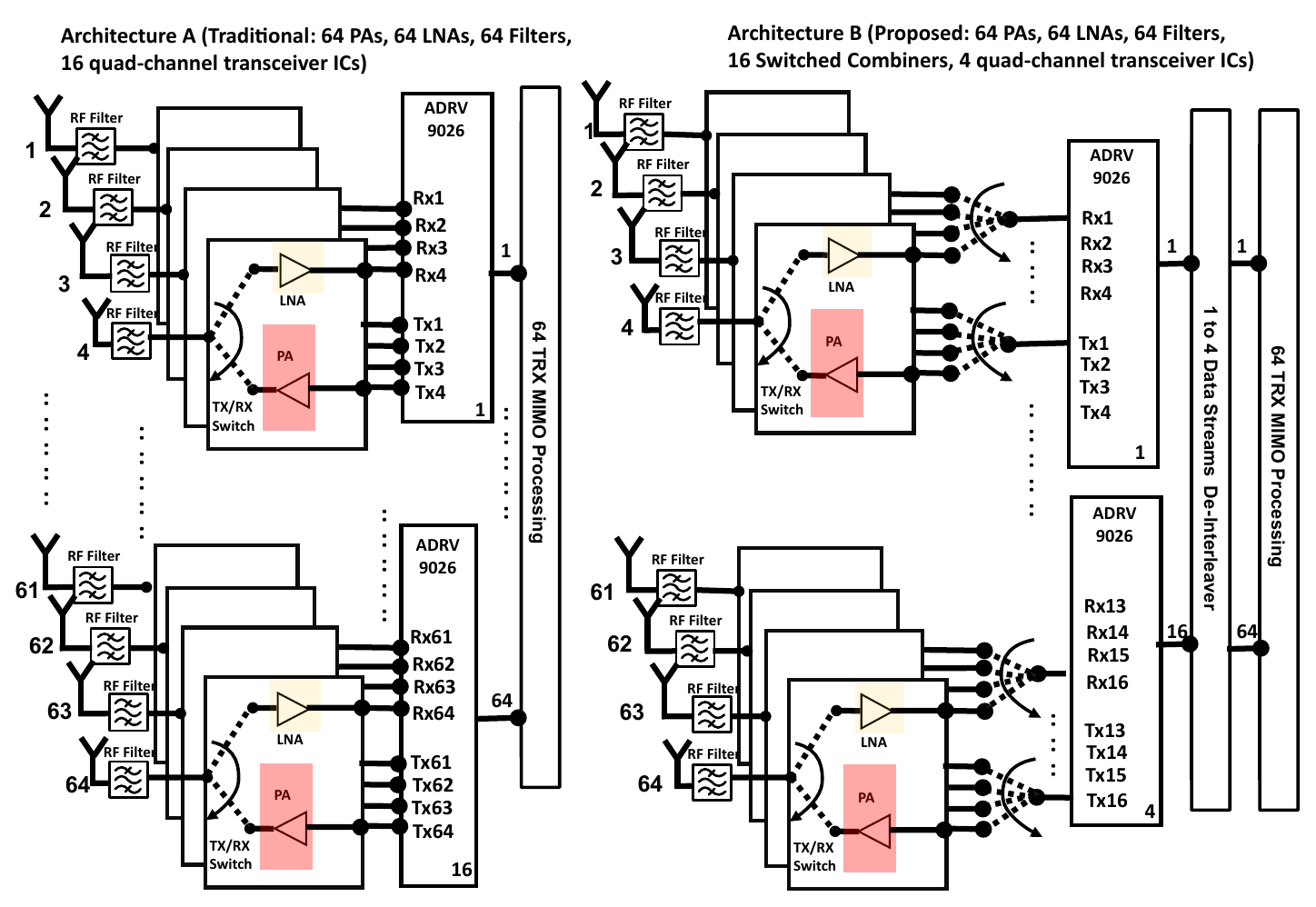}
    \vspace{3pt}
    \caption{Showcasing how the proposed approach in paper can be incorporated to reduce number of transceiver ICs}
    \label{fig:arch}
\end{figure}
In order to simulate the effects of the cross-talk, we force the equalization process to use SISO equalization, by artificially setting the non diagonal entries of the $4\times4$ PUSCH channel matrix per subcarrier as zero. This forces the Matlab implementation of NR equalizer to treat the problem as a SISO problem, since the channel estimation doesn't carry information about the cross-talk anymore. Hence, via this process, we can obtain the three SNRs of interest. that is, (i) SNR of interface (a) using dedicated ADCs, (ii) SNR of interface (b) using single ADC with SISO equalization and (iii) SNR of interface (b) using single ADC with MIMO equalization.

\subsection{Reconstruction SNR Results}

We show the spectrum outputs in Fig. \ref{fig:spectrum}, with cutoff frequency $F_c = 400$ MHz. For all the simulations, ADC $F_s = 1600$ MHz. As explained in the previous Section, for interface (a), ADC $F_s = 400$ MHz, which makes interfaces (a) and (b) achieving similar SNRs, as shown in Fig. \ref{fig:results1} (i) when low pass filter is bypassed. This is the same setting as explored by past work.

However, for Fig. \ref{fig:results1} (ii)-(iii), the low pass filter's cutoff frequency causes some of the higher harmonics to get filtered out. This breaks the orthogonality as discussed in prior Sections. We can see from Fig. \ref{fig:results1} (ii), the cross-talk level is about $20$ dB lower, since the SNR saturates at $20$ dB, and before that it follows the ideal curve. That is, the cross talks become an issue only after $15$ dB, since before that the SNR is noise limited and not cross talk limited. 
As the low pass filter becomes steeper, we can see from Fig. \ref{fig:results1} (iii) that the cross talk level is now $15$ dB lower, as SNR flattens at $15$ dB, with the curve before $10$ dB following the ideal curve. 
In both (ii), (iii) MIMO equalization rcovers the SNR back to the ideal level.

But, as the low pass filter gets steeper, with cutoff frequency going below $400$ MHz, it causes degradation in SNR, since now the effective matrix is having substantial losses. This shows up as about a $7.5$ dB degradation in SNR, since a large part of the harmonics are filtered out. However, this is a corner case, it is not expected for systems too operate at this point, since we are interfacing $4*100$ MHz worth of digital data, and hence, the low pass filtering needs to be commensurate and respect the info-theoretic constraint. We add Fig. \ref{fig:results1} to complete the treatment and show all possible cases of operation.

\subsection{Improvement in Procurement Cost}

So far, we have shown that a single ADc, clocked at $1600$ MHz to receive $4$ 100 MHz streams, achieved the same SNR as $4$ ADCs, clocked at $400$ MHz to receive a $100$ MHz signal each. 
This requires extra hardware in form of a $4\to 1$, `switched-combiner' interface, which utilizes switches with nanosecond-switching capability, such as QS4A210 \cite{qs4a210}, which costs $5\$$ per unit.
\begin{wrapfigure}{r}{0.3\textwidth}
  \begin{center}
    \includegraphics[width=0.95\linewidth]{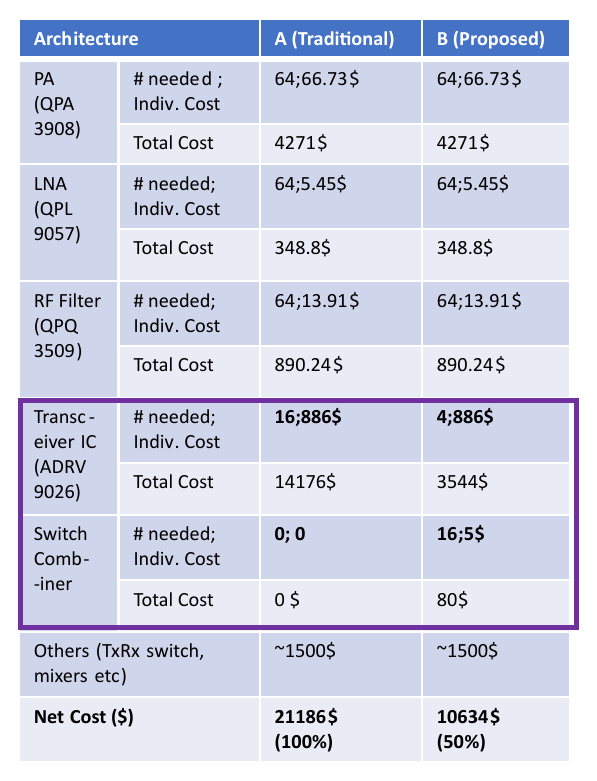}
  \end{center}
  \caption{Cost comparison between the traditional and proposed architectures}
  \label{fig:table_cost}
\end{wrapfigure}
However this leads to a reduction in number of transceiver ICs, that can overall reduce the cost of the MIMO radio. 
In order to quantify the reduction in cost, we take example of building a $64$ antenna Massive MIMO radio in C-band, using a popular transciever IC, ADRV9026 which supports $4$ independent ADC/DAC channels.
We build such a $64$ antenna radio using 16 such ADRV9026 (Architecture A), and compare it to our approach of building it with just $4$ transcivers, with each transciever supporting $4x$ streams loaded into the wider bandwidth, via the switched-combiner interface (Architecture B).
This is visually illustrated in Fig. \ref{fig:arch}.
We use the components from a leading electronics supplier to 5G base-stations, Qorvo, to quantify the various costs required for C-band PA, LNA and RF filter \cite{qorvo}. 
We observe that Architecture B is $50$\% lower cost than Architecture A, as shown in Table \ref{fig:table_cost}.
The cost assumed here is the list price for all items, not including bulk ordering discounts.

\section{Conclusion and Future Work}
In this paper, we showed how RF sampling ADCs, which can sample at $F_s$, much faster than the signal bandwidth $B$, i.e. $F_s>>B$, can be used to accommodate multiple antenna streams. This required extra switch-combiner interface which need to be clocked at orthogonal time clocks. 
We showed how this orthogonality breaks down due to ADC decimation filters, and how MIMO equalization is able to mitigate this, and still preserve the reconstruction SNRs to that of traditional implementations.
Since the proposed scheme reduced the number of digital transceivers, this leads to a cost reduction of about $50\%$ which can lead to lower cost Massive MIMO radios for greater adoption in 6G and onwards.

Future work in this direction can include a theoretical derivation of the exact cross-talk matrix depending on filter cutoff frequency, generalization of the shared ADC interface to a shared DAC interface, and exploring different orthogonalization techniques for analog code designs.

\bibliographystyle{unsrt}
\bibliography{main}
\end{document}